\documentclass[%
reprint, 
superscriptaddress,
amsmath,amssymb,
aps,
pra,
floatfix,
]{revtex4-2}

\usepackage[T1]{fontenc}
\usepackage[english]{babel}
\usepackage{graphicx}
\usepackage{dcolumn}
\usepackage{bm}
\usepackage{siunitx}
\usepackage{todonotes}
\usepackage{physics}
\newcommand{\micron}{\micro\meter}
\usepackage{blindtext}
\usepackage{hyperref}
\usepackage{tablefootnote}
\usepackage[capitalise]{cleveref}
\DeclareUnicodeCharacter{2212}{-}
\newcommand{\cryopaper}{\cite{fontanaMechanicallyStableTunable2021}}
\newcommand{\gevpaper}{\cite{hoyjensenCavityEnhancedPhotonEmission2020}}
\newcommand{\ghzpi}[2]{$#1/2\pi = \SI{#2}{\giga\hertz}$}

\begin{document}

\preprint{APS/123-QED}

\title{Lifetime Reduction of Single Germanium-Vacancy Centers in Diamond via a Tunable Open Microcavity}
\author{Rigel Zifkin}
\thanks{These authors contributed equally to this work}
\affiliation{Department of Physics, McGill University, 3600 Rue University, Montreal Quebec H3A 2T8, Canada}

\author{C\'esar Daniel Rodr\'iguez Rosenblueth}
\thanks{These authors contributed equally to this work}
\affiliation{Department of Physics, McGill University, 3600 Rue University, Montreal Quebec H3A 2T8, Canada}

\author{Erika Janitz}
\affiliation{Department of Electrical and Software Engineering, University of Calgary, 2500 University Drive NW, Calgary Alberta T2N 1N4, Canada}
\author{Yannik Fontana}
\affiliation{Department of Physics, University of Basel, Klingelbergstrasse 82, 4056 Basel, Switzerland}
\author{Lilian Childress}%
\email{lilian.childress@mcgill.ca}
\affiliation{Department of Physics, McGill University, 3600 Rue University, Montreal Quebec H3A 2T8, Canada}

\date{\today}

\begin{abstract}
Coupling between a single quantum emitter and an optical cavity presents a key capability for future quantum networking applications. Here, we explore interactions between individual germanium-vacancy (GeV) defects in diamond and an open microcavity at cryogenic temperatures. Exploiting the tunability of our microcavity system to characterize and select emitters, we observe a Purcell-effect-induced lifetime reduction of up to $4.5\pm0.3$, and extract coherent coupling rates up to $360\pm20$ MHz. Our results indicate that the GeV defect has favorable optical properties for cavity coupling, with a quantum efficiency of at least $0.34\pm0.05$ and likely much higher.  
\end{abstract}

\maketitle


\section{Introduction}
\label{sec:intro}
Combining long-lived electronic and nuclear spin states with atomic-like optical transitions, defect centers in diamond offer a promising platform for quantum networking applications.  
The most well-established defect, the nitrogen-vacancy (NV) center, combines access to nuclear spins possessing extraordinarily long spin coherence times~\cite{maurerRoomTemperatureQuantum2012, bradleyTenQubitSolidStateSpin2019} with spin-photon entanglement~\cite{toganQuantumEntanglement2010}, enabling pioneering demonstrations of multi-node quantum networks
~\cite{pompiliRealizationMultinodeQuantum2021}. 
To realize the potential of this platform, however, requires mitigation of the NV's low zero-phonon-line (ZPL) emission (3\%) and its spectral diffusion, which significantly hinder its efficiency as a spin-photon interface.  

The group-IV-vacancy defects in diamond have been recently explored as alternatives to the NV with superior optical properties~\cite{bradacQuantumNanophotonics2019}. The silicon vacancy (SiV), germanium vacancy (GeV) and tin vacancy (SnV) 
emit over 60\% of their fluorescence into the ZPL
~\cite{janitzCavityQuantumElectrodynamics2020}, and
their symmetry reduces susceptibility to fluctuating electric fields, inhibiting spectral diffusion. While dilution refrigerator temperatures are needed to observe long spin coherence times in bulk SiV~\cite{sukachevSiliconVacancySpin2017}, the heavier group-IV elements and strain engineering ~\cite{sohnControllingCoherenceDiamond2018, assumpcaoDeterministicCreationStrained2023} could enable 4K operation~\cite{guoMicrowavebasedQuantum2023}.  

Improved coherent photon emission and collection rates can also be achieved by coupling defect centers to optical resonators, leveraging the Purcell effect.
Group-IV defects have been successfully incorporated into nano-photonic resonators with high cooperativity ($C\gg1$), enabling demonstration of memory-enhanced quantum communication using an SiV \cite{bhaskarExperimentalDemonstrationMemoryenhanced2020}. 
Open-geometry Fabry-P\'erot microcavities offer an alternate high-performance approach that can in principle achieve comparable coupling rates while using minimally fabricated diamond membranes that preserve the bulk optical properties of the embedded emitters~\cite{rufOpticallyCoherentNitrogenVacancy2019}, and facilitate membrane-based strain engineering~\cite{guoMicrowavebasedQuantum2023}. 
Additionally, the tunability of open cavities permits {\it in-situ} control over cavity frequency as well as selection of emitters with optimal properties, 
providing a more flexible experimental platform. 
Such cavities, at cryogenic temperatures, have shown significant Purcell enhancement of NV centers in bulk-like membranes \cite{riedelDeterministicEnhancementCoherent2017,rufResonantExcitationPurcell2021,yurgensCavityassistedResonance2024}, SiVs in membranes \cite{benedikterCavityEnhancedSinglePhoton2017, hausslerDiamondPhotonics2019} as well as nanodiamonds \cite{bayerOpticalDrivingSpin2023}, and membrane-based SnVs, where cooperativity approaching 1 has very recently been observed~\cite{herrmannCoherentCoupling2023}.

Room-temperature GeV experiments in open cavities \cite{hoyjensenCavityEnhancedPhotonEmission2020,feuchtmayrEnhancedSpectral2023} and plasmonic resonators \cite{kumarFluorescenceEnhancement2021}, as well as a cryogenic nanophotonic waveguide \cite{bhaskarQuantumNonlinearOptics2017} have indicated that GeVs possess a high quantum efficiency, making them favorable candidates for strong cavity coupling and Purcell enhancement.
In addition, GeVs can be readily created with standard implantation and vacuum annealing techniques \cite{iwasakiGermaniumVacancySingle2015,bhaskarQuantumNonlinearOptics2017}, or incorporated during growth via chemical vapor deposition \cite{sedovGrowthPolycrystalline2018}. Furthermore, their moderate spin-orbit splitting facilitates excellent spin coherence times at dilution refrigerator temperatures ($\sim20$ ms \cite{senkallaGermaniumVacancy2024}), while intermediate strain-splitting should enable straightforward microwave spin control \cite{sukachevSiliconVacancySpin2017,pingaultCoherentControl2017,rosenthalMicrowaveSpin2023,guoMicrowavebasedQuantum2023}. This combination of characteristics motivates further investigation of GeV defects for quantum networking applications.


Here, 
we study the coupling of single GeVs in a diamond membrane to a Fabry-P\'erot microcavity at cryogenic temperatures, observing a significant reduction of the excited-state lifetime. The lifetime modification is larger than what has been reported to date for diamond defects in open microcavities, illustrating the favorable ZPL fraction and quantum efficiency of GeVs, as well as 
the stable operation of a high-finesse resonator. 
We also explore the capabilities of open cavities for {\it in-situ} characterization and selection of optimal emitters, exploiting the narrow-linewidth cavity resonances to extract emitter fine-structure splittings and dephasing rates. We observe a broad range of ground-state splittings, some of which exceed bulk values by more than a factor of 6; such large values are promising for future strain engineering of spin states by selection of strain-split outliers. After accounting for the impact of cavity vibrations, we extract an on-resonance emitter-cavity coupling rate of $360\pm20$ MHz, allowing us to place a lower bound on the quantum efficiency of the GeV of $0.34\pm0.05$. Moreover, with reduced emitter dephasing, the observed coupling would be sufficient to achieve cooperativity $C > 1$. 



\section{Experimental Setup}
\label{sec:exp_setup}

\begin{figure*}[ht]
    \includegraphics{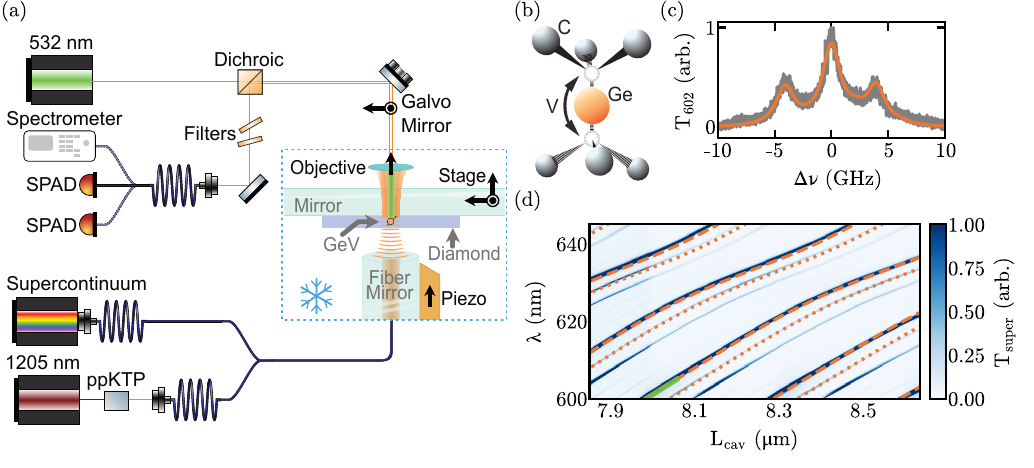}
    \caption{ 
    Experimental overview. (a) Setup Schematic: Blue dashed box indicates the cryogenic system, containing the fiber cavity, sample and imaging objective. 
    Excitation (532 nm) and filtered fluorescence collection is performed in free space (upper path). 
    Collected light is optionally directed to a single photon avalanche detector (SPAD), a 50/50 beam splitter followed by two SPADs, or a spectrometer.
   Cavity characterization measurements employ fiber-coupled supercontinuum or $\SI{602}{\nano\meter}$ laser light produced by second-harmonic generation in a ppKTP waveguide (lower path). 
    Black arrows indicate motional degrees of freedom.
    (b) Model of GeV structure showing nearby carbon atoms (C), interstitial germanium atom (Ge), and associated split-vacancy (V).
    (c) Cavity scan over resonance at $\sim\SI{602}{\nano\meter}$ with 4 GHz sidebands. 
    Solid line indicates fit yielding the cavity linewidth of $\SI{2.07\pm0.06}{\giga\hertz}$.
    (d) Cavity transmission probed by the supercontinuum source over a range of cavity lengths. 
    Dashed and dotted lines indicate fits for two transverse modes to a numerical model \gevpaper, from which the cavity geometry is determined. 
    The green-highlighted portion is used to evaluate the slope of the resonance employed in our experiments.}
    \label{fig:setup}
\end{figure*}

The experimental platform, shown in \cref{fig:setup}a, consists of an optical cavity formed by a laser-machined fiber 
mirror and a macroscopic planar mirror to which the 
diamond sample is bonded. 
The planar mirror is mounted on a three-axis positioning stage that enables both long-range (mm) motion and fine (nm) tuning of the emitter location within the cavity mode. 
Additionally, the fiber mirror is mounted on a shear piezo for fast adjustment of the cavity length. 
The apparatus is placed on a custom vibration-isolation platform mounted in a closed-cycle cryostat with high-numerical-aperture optical access, permitting spatially selective excitation of emitters within the cavity mode~\cryopaper. 

The diamond sample is produced from an electronic grade diamond implanted with germanium ions.
Following the process outlined in Appendix \ref{sec:appendix_sample_fabrication}, we fabricate a $\sim \SI{1}{um}$ thick diamond membrane with a density of $<1$ GeV/\SI{}{/\micron\squared} and bond it to the mirror via intermolecular forces \cite{hoyjensenCavityEnhancedPhotonEmission2020}.

The vibrations from the cryocooler pose a significant challenge to stable cavity operation, causing the mirror position to be detuned by $\SI{51\pm1}{\pico\meter}$ RMS in a 10 kHz bandwidth. However, this is reduced to $\SI{28.9\pm0.7}{\pico\meter}$ during the quiet period between each actuation of the cryocooler, providing a $\sim33\%$ low-noise duty cycle (see \cref{sec:appendix_vibration}). 
The RMS motion measured here is larger than that previously measured for this apparatus \cryopaper, which we attribute to an increase in noise from amplifiers used for the control of piezo stages and shifting in the system from several cooldown cycles.


The optical cavity's geometric parameters are characterized by transmission white-light spectroscopy performed at varying cavity lengths (see \cref{fig:setup}d)~\cite{janitzFabryPerotMicrocavityDiamondbased2015}. By fitting a numerical model to the resulting spectrograph, we determine the fiber-mirror radius of curvature 
to be $\SI{22.5 \pm 0.3}{\micron}$ and the diamond thickness 
to be $\SI{1.05 \pm 0.02}{\micron}$. At our measured diamond thickness, the emission wavelength of GeVs will couple to air-like cavity modes~\gevpaper.
We also fit the white-light spectrograph for the slope of the resonance in the vicinity of the emission wavelength, yielding a value of $\SI{17.64\pm0.01}{\pico\meter/\giga\hertz}$, allowing for the conversion of linewidths between frequency and length.

To determine the cavity linewidth, we probe the cavity with a laser near the GeV ZPL wavelength of \SI{602}{\nano\meter}. 
The beam is phase modulated by an electro-optic-modulator, providing sidebands at $\pm\SI{4}{\giga\hertz}$. 
Sweeping the cavity length $L$ through the resonance near $L=\SI{8.00\pm0.05}{\micron}$, and triggering data acquisition to occur within the cryostat quiet period, the known sideband spacing allows us to determine the cavity linewidth to be $\SI{2.07\pm0.06}{\giga\hertz}$, corresponding to a length linewidth of $\SI{37\pm1}{\pico\meter}$, and a finesse $\sim8250$ (see \cref{fig:setup}c).

\begin{figure*}[ht]
    \includegraphics{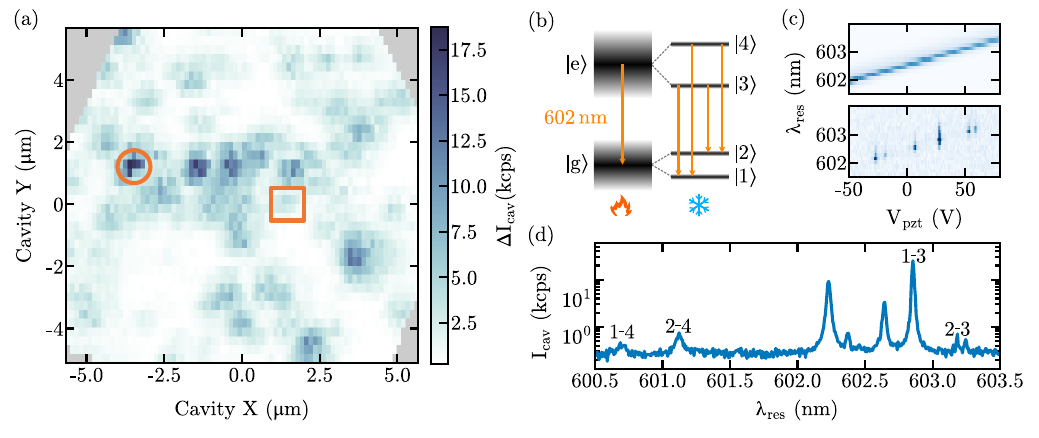}

    \caption{Cavity scans. (a) 2D map of cavity coupled emission amplitude (max-min). Orange circle indicates emitter studied in detail throughout the main text. Orange square indicates additional emitter for which cQED parameters are determined.
    Gray shaded area delimits the range of the piezo stage.
    (b) Level structure diagram of the GeV showing how the single broadened \SI{602}{\nano\meter} optical transition at room temperature splits  at cryogenic temperatures into the four narrow lines identified in (d).
    (c) Calibration of axial cavity scan. (Top) Cavity transmission spectrum vs fiber piezo voltage and (bottom) scan over the same range while exciting the emitter with CW \SI{532}{\nano\meter} laser light. Color scales for both data sets are arbitrary.
    (d) Axial cavity scan at circled position in (a) (logarithmic scale). Peaks belonging to the studied emitter are labelled by transitions between the levels shown in (b).
    }
    \label{fig:3dscan}
\end{figure*}

To locate individual emitters, the sample is pumped with continuous-wave \SI{532}{\nano\meter} laser light, while the positioning stage scans the sample relative to the fiber mirror. Acquisition occurs during the entire cryostat pump cycle. 
Light emitted from the cavity is collected via the macroscopic flat mirror in free space, separated from the excitation beam by a dichroic mirror and additionally filtered to the range of $\sim600-605$ nm for the rejection of background fluorescence.
At each lateral position of the sample, the fiber piezo sweeps the cavity length, tuning the resonant wavelength around \SI{602}{\nano\meter}; we quantify the cavity-coupled emission by the difference between the maximum and minimum fluorescence observed during the cavity-length scan $(\Delta \rm{I}_{\rm{cav}})$. 
\cref{fig:3dscan}a shows $\Delta \rm{I}_{\rm{cav}}$ over a ~$(\SI{10}{\micro\meter})^2$ lateral scan of our sample. Note that the fiber piezo's short range of motion at cryogenic temperatures impedes large-area scans, as small tilts of the sample and mirror shift the cavity resonant wavelength outside the collected range. 
To circumvent this limit, we correct for tilt by adjusting the piezo stage height in proportion to its lateral displacement.

\section{Emitter Characterization}
\label{sec:emitter_charac}

Over the area scanned in \cref{fig:3dscan}a, we see several point-like sources of cavity-coupled fluorescence, indicating the presence of localized emitters. Examining the spectra of the emitted fluorescence -- encoded in the axial cavity scans -- allows us to identify them as GeV defects, which characteristically emit in four ZPL lines near \SI{602}{\nano\meter} due to splitting of the ground and excited states (see \cref{fig:3dscan}b) \cite{bhaskarQuantumNonlinearOptics2017}.  
We calibrate the axial scans by performing cavity-length sweeps while recording fluorescence on a spectrometer, shown in \cref{fig:3dscan}c-d; the cavity is first excited with a supercontinuum laser, giving a linear relationship between piezo voltage and resonant wavelength, and secondly, we excite the emitter (indicated by a circle in \cref{fig:3dscan}a) at 532 nm, yielding its spectrum.
Note that the features in the lower part of \cref{fig:3dscan}c are vertically elongated, indicating superior resolution of the axial cavity scan relative to the spectrometer.
We identify the four emission lines corresponding to the centered GeV (\cref{fig:3dscan}d) by comparing spectra at different lateral positions.
Approaching the GeV of interest, the four labelled peaks simultaneously reach their maxima, while the other peak heights vary monotonically.
We therefore associate the other peaks in the spectra with nearby GeVs.

\begin{figure}[ht]
    \centering
    \includegraphics{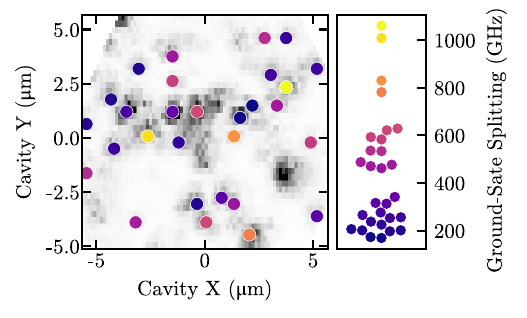}
    \caption{Spectral analysis. (Left) Duplicate of scan shown in \cref{fig:3dscan}a. Each point represents spectral analysis of an emitter, with ground-state splitting (extracted from 1-3/2-3 peak splitting) indicated by color.  
    (Right) Definition of color scale and distribution of splittings from left-hand scan. X-axis is arbitrary and set to minimize point overlap.}
    \label{fig:hyperspectral}
\end{figure}

Beyond identifying the emitters as GeVs, the spectral resolution of the scanning cavity system permits mapping out the energy splitting between the transitions of individual emitters.
In \cref{fig:hyperspectral} we map out the ground-state orbital splitting (i.e. the 1-3/2-3 peak splitting) over several GeVs present in the analyzed region. 
We observe two well-populated groupings in the splitting distribution with a small number of outliers, which could be explained by uniform strain in the sample differently affecting the four possible orientations of GeVs. Moreover, most emitters exhibit a larger splitting than the \SI{152}{\giga\hertz} value reported in Ref.~\cite{bhaskarQuantumNonlinearOptics2017}, indicating a high-strain environment \cite{jahnkeElectronPhonon2015}; some even exceed 1 THz, larger than the unstrained value for the SnV.
While our device was not fabricated with the intent of introducing large strain, our results are comparable to Ref.~\cite{guoMicrowavebasedQuantum2023}, in which a diamond membrane is strained by differential thermal contraction with a silica substrate, similar to the macroscopic mirror to which our sample is bonded.
Their technique provides an engineerable way to strain the emitters without significantly affecting their optical properties.
Our findings indicate that the intermolecular forces bonding the membrane to the mirror provide enough adhesion to achieve a similar effect.
Such large ground-state splittings could suppress phonon-mediated decoherence of the GeV spin, as has been observed in other group-IV defect systems~\cite{meesalaStrainEngineering2018,sohnControllingCoherenceDiamond2018}, enabling operation at higher temperatures. Moreover, a high-strain environment simplifies microwave spin control, which becomes increasingly challenging for heavier group-IV defects~\cite{guoMicrowavebasedQuantum2023}. The compatibility of the membrane-in-cavity geometry with high strain is thus highly advantageous, and opens the opportunity for further engineering of the strain environment by selection of substrate and deposition processes~\cite{guoMicrowavebasedQuantum2023}.

\begin{figure*}[ht]
    \includegraphics{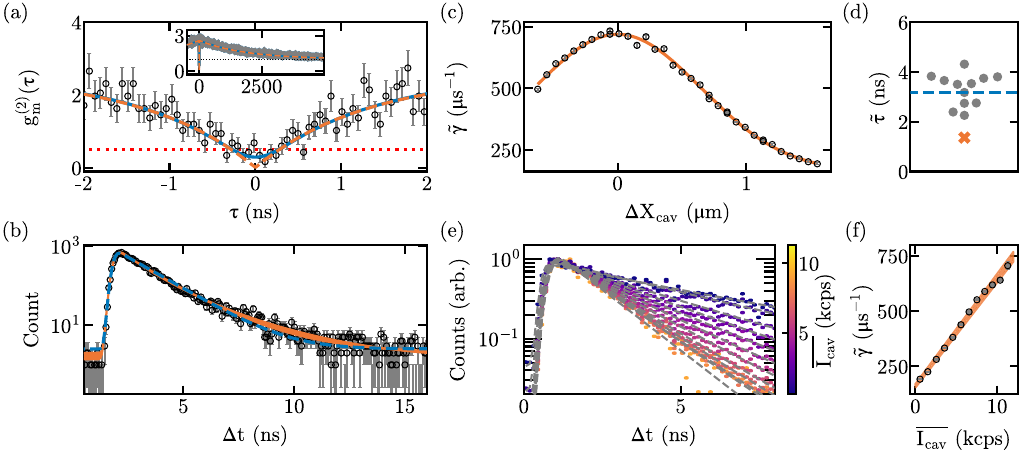}
    \caption{Cavity-modified emitter lifetime. (a) $g^{(2)}_m(\tau)$ acquired with CW excitation. 
    Solid (dashed) line shows fit with (without) convolution with the detector response function, yielding $g^{(2)}_m(0) = 0.29\pm0.07\;(0.0094\pm0.0004)$.
    Inset: large-time behaviour with fits asymptotically approaching 1 (dotted black line).  (b) TCSPC measurement of GeV fluorescence under pulsed 531 nm excitation. Dashed blue line shows fit to an exponentially modified Gaussian function, yielding a lifetime of \SI{1.36\pm0.01}{\nano\second}. Shaded orange region shows fit to numerical model, see text for details. (c) Decay rate as a function of emitter lateral displacement relative to the cavity mode. Orange line shows Gaussian fit. Error bars from lifetime fit uncertainties are smaller than the data points.
    (d) Distribution of lifetimes measured for emitters shown in \cref{fig:3dscan}a. Placement on X-axis is arbitrary to avoid overlap of points. Dashed blue line shows sample mean. Orange 'x' indicates lifetime of emitter studied in detail. 
    (e) Normalized time-tagged TCSPC measurements filtered by mean emission intensity ($\overline{I_{\rm{cav}}}$). Gray dashed lines correspond to Gaussian modified exponential fits. 
    (f) Decay rates ($\tilde{\gamma}$) computed from fit in (e) as a function of mean emission intensity. Shaded region shows the 1-sigma confidence interval of a linear, orthogonal-distance-regression fit. 
    }
    \label{fig:lifetime}
\end{figure*}

Here, we take advantage of the tunability of our system to select a particularly bright, isolated emitter (indicated by a circle in \cref{fig:3dscan}a) for further investigation.
To confirm that the emission is from a single GeV, we perform a Hanbury-Brown-Twiss measurement on the collected fluorescence from the bright 1-3 transition. 
The collected and filtered fluorescence is sent to a 50/50 fiber beam splitter with each output coupled to a single-photon avalanche diode (SPAD). 
To account for cavity-length drift and vibrations from the full cryostat cycle, we post-select for coincidence events within each \SI{10}{\milli\second} interval containing more than 50 counts. 
\Cref{fig:lifetime}a shows the normalized delay time distribution.
To determine whether our data is consistent with a single-photon emission, we fit it with a background-corrected three-level bunching function convolved with a Gaussian instrument response,
\begin{align} 
g_m^{(2)}(\tau) = \left(Ne^{-\frac{1}{2}\left({\tau}/{\sigma_t}\right)^2}\right)\ast\left(g^{(2)}(\tau)s^2 + 1 - s^2\right).
\end{align}
Here $x(t) \ast y(t)$ indicates $(x\ast y) (t)$, $\sigma_t$ is the response time, $N$ is a normalization constant, $s$ is the fraction of signal counts coming from the emitter, and 
\begin{align}g^{(2)}(\tau) = 1 - (1 + a)e^{-\frac{|\tau|}{\tau_1}} + ae^{-\frac{|\tau|}{\tau_2}},
\end{align}
where $\tau_1$ and $\tau_2$ are the lifetimes of the excited state and shelving state, respectively. The latter is associated with the prominent bunching observed and believed to arise from a dark charge state of the GeV \cite{neuPhotophysicsSingle2012, chenOpticalGating2019, hoyjensenCavityEnhancedPhotonEmission2020}.
The convolution is necessary, as the fitted response time $\sigma_t = \SI{196}{\pico\second}$ masks the sharp rise-time $\tau_1 = \SI{1.31\pm0.06}{\nano\second}$, resulting in the lowest value of $g^{(2)}_m$ that we observe being $0.2\pm 0.4$. 
By comparing fluorescence counts with the cavity on and off resonance, we independently measure $s = \SI{0.9953\pm0.0002}{}$, and fit for the other parameters. 
We find good agreement between the model and data ($\chi^2_{reduced} =  \SI{1.03\pm0.08}{}$ for data within $\pm10$ ns from $\tau=0$), demonstrating consistency with single-photon emission.

The short lifetime $\tau_1$ in the $g^{(2)}_m$ data is indicative of a fast emitter lifetime, which we confirm through Time-Correlated Single Photon Counting (TCSPC) by pumping the emitter with a pulsed (FWHM < \SI{100}{\pico\second}) 531nm laser and collecting a histogram of delay times between the collected fluorescence and laser pulse. 
To minimize the impact of cavity vibrations on the measured lifetimes, counts are only collected during the low-vibration period between actuations of the cryocooler.
We fit the data with an exponentially modified Gaussian distribution, which captures the exponential decay of the emitter 
and the response time of the instrumentation. As shown in \cref{fig:lifetime}b, for the studied emitter, we observe a minimum lifetime of $\tau_{\rm{min}} = \SI{1.36\pm0.01}{\nano\second}$.

We verify that the lifetime reduction is modified by the GeV's placement relative to the cavity mode by repeating lifetime measurements with the cavity displaced laterally, shown in \cref{fig:lifetime}c.
Fitting the measured decay rates ($\tilde{\gamma}$) versus lateral position to a Gaussian mode profile yields a $1/e^2$ waist of \SI{1.30 \pm 0.01}{\micron}, slightly smaller than the expected value of \SI{1.42}{\micron} calculated from the geometric cavity parameters \cite{damOptimalDesignDiamondair2018}.
From the constant offset of the fit, corresponding to the decay rate at infinite displacement, we compute a bare GeV lifetime of $\tau = \SI{6.1\pm0.1}{\nano\second}$, consistent with the literature value of \SI{6.0\pm0.1}{\nano\second} \cite{bhaskarQuantumNonlinearOptics2017,hoyjensenCavityEnhancedPhotonEmission2020} for free-space emission in bulk diamond.
Therefore, we measure a maximum lifetime reduction factor of $4.5\pm0.3$, indicating that $78\pm8\%$ of the emitted photons are coupled to the cavity.

Slow drifts in the cavity length prevented measuring lifetimes at static detunings.
So, to verify the variation of Purcell enhancement with cavity detuning, we collected time-tagged TCSPC measurements while sinusoidally driving the fiber mirror through resonance at $\SI{1}{\hertz}$.
By post-selecting events based on the mean count rate ($\overline{I_{\rm{cav}}}$) in each \SI{10}{\milli\second} acquisition window, we can determine the lifetime measured for different cavity detunings, see \cref{fig:lifetime}e. 
The resulting values of $\tilde{\gamma}$ follow a linear trend with the mean count rate, as shown in \cref{fig:lifetime}f, demonstrating that the decay rate follows the Lorentzian profile of the collected fluorescence rate. 
As above, fitting for the off-resonant decay rate yields $\tau = \SI{6.2\pm0.3}{\nano\second}$, consistent with the bulk value.

Including the lifetimes of several other emitters in \cref{fig:3dscan}, we find a range of enhanced decay rates, as shown in \cref{fig:lifetime}d, with mean value \SI{3.2\pm0.2}{\nano\second}. However, we note that the alignment to each emitter was not optimized to minimize the measured lifetimes, and each bright spot was not confirmed to be a single GeV. 
These measurements demonstrate a significant Purcell enhancement across all studied emitters.

\section{Estimation of cQED Parameters}
\label{sec:cqed}

The experimental observations detailed above can be interpreted by modelling our cavity quantum electrodynamics (cQED) system as an emitter with decay rate $\gamma$ and pure dephasing rate $\gamma^*$ that interacts with a cavity with decay rate $\kappa$ via a coherent coupling rate $g$ \cite{jaynesComparisonQuantum1963,shoreJaynesCummingsModel1993}. 
For systems in the weak-coupling limit, $g \ll \kappa,\gamma,\gamma^*$, the cavity adds a new decay path for emitter excitations via the cavity field, resulting in a new Purcell-enhanced decay rate 
$\gamma_{\rm{P}} = \left(\tau_{\rm{P}}\right)^{-1} = \gamma + \gamma_{\rm{cav}}$, where $\tau_{\rm{P}}$ is the new excited-state lifetime.
As detailed in \cref{sec:appendix_cqed}, 
the lifetime reduction factor is given by
\begin{equation} \label{eq:tauratio}
    \frac{\tau}{\tau_{\rm{P}}}= 1 + \frac{\gamma_{\rm{cav}}}{\gamma} = 1 + \frac{4 g^2}{\kappa\gamma}\frac{\kappa^2}{4g^2 +\Gamma \kappa}\frac{\Delta^2}{\Delta^2 + 4 \delta^2}
\end{equation}
where $\delta$ is the emitter-cavity detuning, $\Gamma = \gamma+\gamma^*+\kappa$ is the total system decoherence rate, and $\Delta = \sqrt{\Gamma^2 + 4g^2 \Gamma/\kappa}$ is the linewidth in detuning. 
We note that in the fast-cavity limit ($\kappa \gg g,\gamma,\gamma^*$) $\Gamma \approx \kappa$, and the on-resonance expression reduces to the usual definition of the cooperativity, $\gamma_{\rm{cav}}/\gamma = 4 g^2/\kappa \gamma$. 

To determine how close our system is to coherent operation, we seek to extract all of the cQED parameters, notably the coherent coupling rate $g$.
However, we cannot immediately extract $g$ from our observed maximum lifetime reduction factor $\tau/\tau_{\rm{min}}$ for two reasons:
(1) our cavity does not satisfy $\kappa \gg \gamma^*$, so we need to consider the full form of Eq.~\ref{eq:tauratio}, which requires knowledge of $\gamma^*$ and (2) cavity vibrations induce nonzero detuning of our cavity from the emitter resonance, such that $\tau_{\rm{min}} \neq \tau_{\rm{P}} (\delta = 0)$, requiring a more careful analysis to extract the on-resonance coupling rate.

 $\gamma^*$ is typically characterized by resonant photoluminescence excitation experiments. We instead examine the emitter linewidth under continuous-wave (CW) off-resonant (532 nm) excitation by scanning the narrow cavity resonance over the emission spectrum. Unlike external scanning Fabry-P\'erot spectrometers \cite{bernienTwoPhotonQuantum2012}, these intra-cavity spectra must be interpreted within the coupled emitter-cavity model.  
In particular, additional dephasing from the CW incoherent pump rate $\Pi$ increases the emission linewidth to 
\begin{equation} \label{eq:spect_lw}
    \Delta_{\rm{cw}} = \sqrt{\Gamma_{\rm{cw}} \left(\Gamma_{\rm{cw}}+\frac{4 g^2 (\gamma +\Pi +\kappa )}{\kappa (\gamma +\Pi )}\right)}
\end{equation}
where $\Gamma_{\rm{cw}} = (\gamma+\gamma^*+\kappa+\Pi)$ is the total dephasing rate of the system (see \cref{sec:appendix_driven} for details). Thus, while the cavity emission linewidth does not directly reveal $\gamma^*$, when combined with independent measurements of $\kappa, \gamma$ and $\Pi$, it provides a constraint relating $g$ and $\gamma^*$. 

\begin{figure}[h]
    \includegraphics{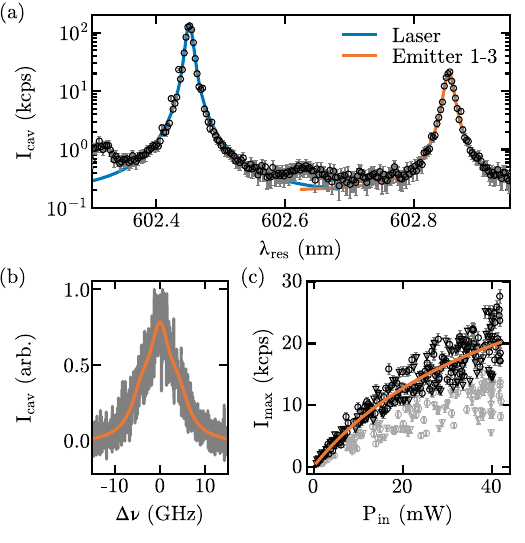}
    \caption{Emitter characterization.
    (a) Narrow-range axial cavity scan over the 1-3 transition of the studied emitter with the \SI{602}{\nano\meter} laser as reference. Solid lines show Voigt-profile fits to each peak. 
    (b) Cavity linewidth characterization as in \cref{fig:setup}c, exhibiting broadening due to damage from overexposure to the high-powered excitation laser, applicable to data shown in (a) only.
    (c) Saturation curve of the studied emitter. 
    Circle and triangle markers indicate two separate acquisitions, with light gray markers being excluded from the combined fit (solid line) as described in the text.}
    \label{fig:emitter}
\end{figure}

To extract $\Delta_{\rm{cw}}$ from experimental spectra, we need to consider cryostat vibrations. Because we acquire these spectra continuously, we must account for the large-scale, Lorentzian-distributed fluctuations that occur during the noisy part of the cryostat cycle, which are not consistent from shot to shot. 
By passing a near-resonant, narrow-linewidth (<MHz) laser through the cavity as the spectrum is acquired, we obtain an additional resonance peak that reveals this vibrational contribution {\it in-situ}. 
\cref{fig:emitter}a shows one such scan, with each peak fit by a Voigt profile constrained to have a fixed Gaussian component commensurate with independent measurements of the Gaussian component of cryostat vibrations. 
We use the Lorentzian components of the Voigt fits to the emitter peak (\ghzpi{\Delta_{\rm{e}}}{11.14\pm0.09}) and laser peak (\ghzpi{\Delta_{\rm{l}}}{8.04\pm0.09}), as well as an updated measurement of cavity linewidth (\ghzpi{\kappa'}{5.8\pm 0.1}, see \cref{fig:emitter}b) to determine the emitter Lorentzian linewidth $\Delta_{\rm{cw}} = \Delta_{\rm{e}} - (\Delta_{\rm{l}} - \kappa') = (8.9 \pm 0.2)\cdot 2\pi$ GHz. The given values result from averaging over 47 scans. 
For further details of this analysis, see \cref{sec:appendix_vibration,sec:appendix_intracav}.

The last unknown parameter required to use Eq.~\ref{eq:spect_lw} is the optical pumping rate $\Pi$. 
To estimate $\Pi$, saturation measurements are performed by recording the peak fluorescence rate at different excitation powers  (see \cref{fig:emitter}c). As saturation data is acquired, we continuously sweep the cavity over the emission line, allowing us to compensate for the shift in peak position with increasing thermal input. 
The large scatter in the data arises from variations in synchronization between cavity scans and the cryostat pump cycle.
To account for this noise, we combine two subsequently acquired data sets and exclude all points below a $5\sigma$ outlier threshold set by an initial fit to the data.
As shown in \cref{fig:emitter}c the portions of removed data are not shared between the runs. 
We fit the combined data with a saturation curve 
\begin{align}
    I(P) = \frac{I_\infty P}{P+P_{\rm{sat}}} + c_{\rm{bg}}P + c_{\rm{dark}},
\end{align}
where $I_\infty$ is the asymptotic maximum count rate, $P_{\rm{sat}}$ the saturation power, and $c_{\rm{bg}}$ and $c_{\rm{dark}}$ are the constants associated with the linear background and dark count corrections, respectively. 
We determine  $c_{\rm{bg}}$ and $c_{\rm{dark}}$ independently by considering the count rates far from resonance in each cavity sweep.
From the fit to \cref{fig:emitter}c, we determine $P_{\rm{sat}} = \SI{52\pm7}{\milli\watt}$ as measured at the top of the cryostat window (of which approximately half reaches the emitters).
The saturation power is elevated compared to GeVs in non-cavity-coupled samples \cite{hoyjensenCavityEnhancedPhotonEmission2020} because the saturation pump-rate ($\Pi_{\rm{sat}}$) depends on the lifetime of the emitter. In particular, on perfect resonance, considering weak coupling and assuming $\gamma \ll \kappa,\gamma^*$,
we find $\Pi_{\rm{sat}} = \tau_P^{-1}$ (see \cref{sec:appendix_driven} for derivation).
Thus, for the 
$P = \SI{35\pm1}{\milli\watt}$ incident power used in spectroscopic measurements of this emitter, 
we estimate $\Pi = 
(P/P_{\rm{sat}})\tau_{\rm{min}}^{-1}$ = $(0.08\pm0.01)\cdot2\pi$ GHz. 
Ultimately, from applying these measurements of $\Delta_{\rm{cw}}$ and $\Pi$ to \cref{eq:spect_lw}, we obtain a relationship between the unknown parameters $g$ and $\gamma^*$, which we can use as a constraint in analyzing lifetime data.

\begin{table*}[th!]
\centering
\begin{tabular}{lcccccc}
     GeV & $\tau_{\rm{min}}$ & $\rm{g}/2\pi$ (GHz) & $\kappa/2\pi$ (GHz) & $\gamma^*/2\pi$ (GHz) & $g_0/2\pi$ (GHz) & Bound on $\eta_{\rm{QE}}$\\
     \hline
\noalign{\vskip 1mm} 
     1 (Circle)~~   & $1.36\pm0.01$ & $0.36\pm0.02$ &$ 2.07\pm0.06$ & $1.0\pm0.9 $ & $~0.80\pm0.04$ & $>0.34\pm0.05$\\
     2 (Square)~~   & $2.27\pm0.02$ & $0.26\pm0.01$ & $2.14\pm0.01$ & $2.7\pm0.8$ & $~0.82\pm0.05$ & $>0.17\pm0.02$\\
     3            & $3.40\pm0.02$ & $0.279\pm0.009$ & $3.50\pm0.05$ & $11\pm1$ & $~0.67\pm0.05$ & $>0.29\pm0.04$
\end{tabular}
\caption{CQED parameters for  three GeVs, where $g_0$ is the ideal cavity coupling parameter. The lower bound on $\eta_{QE}$ is computed by considering only $\eta_{DW}$ (see text for details). GeV 1 and 2 are indicated by the circle and square respectively in~\cref{fig:3dscan}. GeV 3 was from a prior set of data acquired with a cavity length of \SI{10.50\pm0.02}{\micron} and diamond thickness of $\SI{0.79\pm0.02}{\micron}$, and its value of $\Pi$ is estimated from its lifetime.}
\label{tab:cqedparams}
\end{table*}

Despite the detrimental impact of cavity vibrations, the measured lifetimes are remarkably short. As we accumulate photon counts for TCSPC measurements, we preferentially record arrival times when the photon emission rate is high, i.e. when the cavity is near resonance with the emitter. 
The observed lifetimes are thus far shorter than what would be estimated by directly averaging the lifetime reduction factors over the distribution of vibration-induced detunings.
To consider this effect, following a similar derivation as \cite{rufResonantExcitationPurcell2021}, we examine the probability $P(t)$ of detecting a photon at a delay $t$ after the emitter is excited.
We consider the time-dependent expected cavity population at a fixed detuning $\rho_{cc}(t,\delta)$ (see \cref{sec:appendix_cqed} for analytic expressions), weighted by the distribution of cavity detunings.
Since our lifetime data is acquired in sync with the cryostat quiet periods, we employ a Gaussian distribution of cavity detunings with standard deviation \ghzpi{\sigma_\nu}{1.796\pm0.006} found from the quiet-period cavity vibrations converted to an effective frequency detuning (see \cref{sec:appendix_vibration} for details).
Finally, we convolve the result with a Gaussian instrument response function (with associated response time $\sigma_t$ as a fit parameter), obtaining  
\begin{align}
    P(t) \propto \left(\int\rho_{cc}(t,\delta) e^{-\delta^2/(2 \sigma_\nu^2)}\dd\delta\right)\ast \left(e^{{- t^2}/{(2\sigma_t^2)} }\right).
    \label{eq:histogram}
\end{align}

We fit \cref{eq:histogram}, evaluated numerically, to the TCSPC data shown in \cref{fig:lifetime}b, while constraining the relationship between $g$ and $\gamma^*$ according to \cref{eq:spect_lw}.
To account for uncertainties in measured values, the fit is repeated with experimental parameters sampled from their Gaussian distributions. 
The resulting fit functions are illustrated by the shaded orange region in \Cref{fig:lifetime}b, which represents the one-sigma confidence interval. 
The slightly non-exponential behaviour of the model fits our data better than a pure exponential decay, and we extract fit parameters of \ghzpi{g}{0.36\pm0.02} and \ghzpi{\gamma^*}{1.0\pm0.9}. 
These values correspond to maximum cooperativities at zero detuning with no vibrations of $C_{\rm{inc}} = 4 g^2/(\kappa \gamma) =9\pm1$ (incoherent) and $C = \frac{4g^2}{\kappa(\gamma+\gamma^*)} = 0.2\pm0.2$ (coherent).

A similar analysis was performed on two additional emitters, one indicated by a square in \cref{fig:3dscan}a, the other from prior data.
The results are presented in \cref{tab:cqedparams}. 
While the three emitters all exhibit significant cavity coupling, they also have dephasing rates in excess of the lifetime limit.
One explanation is temperature: our sample mount is thermalized at $T=\SI{18\pm1}{\kelvin}$, and temperature-dependent measurements of GeV linewidths suggest  $\gamma^*/2\pi\sim\SI{2}{\giga\hertz}$ at $\SI{20}{\kelvin}$~\cite{chenUltralowPowerCryogenicThermometry2023}, in line with the measured values for the first two emitters. The broader linewidth observed in the third emitter may indicate other mechanisms, though we continue to observe Lorentzian broadening typical of homogeneous processes.
Nevertheless, the first emitter clearly stands out for its shorter lifetime and stronger coupling, which
may arise from a different local environment affecting its dipole moment, 
branching ratio, and/or other properties. Indeed, the opportunity to readily select such promising emitters is an advantage of tunable open-cavity systems over predefined geometries.


\section{Discussion}
\label{sec:discussion}

The coherent coupling rate $g$ extracted from our data can be compared to what might be expected from independently measured parameters. For an ideal, optimally oriented and located emitter, the coherent coupling rate can be predicted from the cavity geometry and the emitter free-space lifetime. This ideal coupling parameter is given by $g_0 = |E_{\rm{max}}||\mu|/\hbar$, where $E_{\rm{max}}$  
 is the maximum value of the single-photon electric field inside of the diamond membrane 
 (see \cref{sec:appendix_estimatingQE})~\cite{riedelCavityEnhancedRamanScattering2020,damOptimalDesignDiamondair2018}; $|\mu|$ is the dipole moment, which can be related to the lifetime for an ideal two-level system. 
 \cref{tab:cqedparams} gives the calculated $g_0$ for 
 each of the studied emitters, using a bare lifetime of $\tau = \SI{6.1\pm0.1}{\nano\second}$. 

As expected, the measured $g$ are smaller than $g_0$ owing to non-idealities of the GeV. These non-idealities can by quantified by $g = \sqrt{\eta} g_0$, where the factor 
\begin{align}
\label{eq:eta}
\eta = \eta_{\rm{QE}}\, \eta_{\rm{DW}} \,\eta_{\rm{BR}}\, \eta_{\rm{Z}} \cos^2\alpha\end{align} 
accounts for the reduction in dipole moment due to sub-unity quantum efficiency ($\eta_{\rm{QE}}$), Debye-Waller factor ($\eta_{\rm{DW}}$), and branching ratio for the cavity-coupled transition ($\eta_{\rm{BR}}$), as well as the reduction in the component of the electric field along the dipole axis due to misplacement of the emitter from a cavity mode maximum ($\eta_{\rm{Z}}$) and misalignment of the dipole and cavity electric field by angle $\alpha$. While the Debye-Waller factor $\eta_{\rm{DW}} \approx 0.6$ has been measured \cite{siyushevOpticalMicrowaveControl2017}, the other values are poorly known. 

Nevertheless, we can use a comparison of $g$ and $g_0$ to place a lower bound on the quantum efficiency of the GeV. Only considering the reported value of $\eta_{\rm{DW}} = 0.6$, the circled GeV places a lower bound of $\eta_{\rm{QE}} > 0.34\pm0.05$ (see \cref{tab:cqedparams} for others), supporting earlier indications of a high GeV quantum efficiency \cite{bhaskarQuantumNonlinearOptics2017}. The three results in Table I differ because they can be impacted to varying degrees by emitter misalignment and position, and it is possible that the branching ratio and quantum efficiency could be affected by local strain.

The lower bounds on $\eta_{\rm{QE}}$ presented in Table I are highly conservative, as they neglect several sources of inefficiencies (outlined in \cref{eq:eta}) which are not precisely known.
To get some sense of a likely value for $\eta_{\rm{QE}}$, we can estimate the other factors in \cref{eq:eta} by making reasonable assumptions given known values for the SiV (see \cref{sec:appendix_estimatingQE} for details).
If we assume that the GeV dipole is oriented predominately along the $\langle111\rangle$ direction \cite{nguyenPhotodynamicsQuantum2019}, as is the case for SiVs \cite{heppElectronicStructure2014,rogersElectronicStructure2014}, then $\cos^2{\alpha} = 2/3$.
Similarly, we use the SiV value for $\eta_{\rm{BR}} = 0.77\pm0.08$ \cite{mullerOpticalSignatures2014}, as it agrees with the relative intensities of the 1-3 and 2-3 spectral peaks (see \cref{fig:3dscan}(d)), which fall in the range of $0.7-0.9$ for a majority of emitters studied in \cref{fig:hyperspectral}. (Such peak intensities cannot themselves provide a precise measurement due to possibly emission-branch-dependent cavity coupling.)
With these values, our measurements indicate quantum efficiencies in excess of $0.66 \pm 0.12, 0.33 \pm 0.06$ and $0.56 \pm 0.10$ for GeVs 1-3, respectively. Finally, while it is likely that we select bright, low-straggle emitters for study, if we add in straggle at the 1-$\sigma$ level 
(see \cref{sec:appendix_estimatingQE}),
our measurements cannot rule out the possibility of unity quantum efficiency.

\section{Conclusion}
\label{sec:conclusion}



Cavity-coupled diamond defect centers are competitive candidates for realization of a coherent and efficient spin-photon interface. While cavity-coupling of NV, SiV, and SnV defects has received considerable attention, only a few studies have investigated single GeV-cavity systems~\cite{hoyjensenCavityEnhancedPhotonEmission2020, kumarFluorescenceEnhancement2021, feuchtmayrEnhancedSpectral2023}, thus far under ambient conditions. Working with single GeV defects within an open microcavity at cryogenic temperatures, our experiments demonstrate favorable properties for this group-IV defect. We observe a large lifetime reduction, to our knowledge the largest observed for diamond defects in open microcavities, indicating a high quantum efficiency. 

One constraint in our experiments was the poor thermalization of our apparatus near $18\pm 1$ K, leading to emitter dephasing far exceeding the radiative emission rate. This constraint inspired development of an intra-cavity spectroscopy technique to probe emitter linewidths without need for a widely-tunable resonant laser, creating possibilities for future studies examining cavity-coupling of spectrally broad defects or employing high Q-factor cavities. 
Such spectral analysis also revealed large ground-state splittings in certain emitters, which portend improved spin properties and higher-temperature operation.
Combined with the ability to image laterally and select emitters with desired dephasing and/or strain properties, this intra-cavity spectroscopy highlights the utility of an open, tunable platform for the \textit{in situ} optimization of a spin-photon interface.

Our results also suggest that highly coherent coupling between GeV defects and open microcavities is attainable. At lower temperatures, GeV linewidths only 2-3 times the radiative limit can be observed~\cite{siyushevOpticalMicrowaveControl2017, bhaskarQuantumNonlinearOptics2017}. 
For the coupling strengths and cavity decay rates observed here, such emitter linewidths would enable cooperativities in the range of 3-4, 
in the strong-coupling regime.
Additionally, refined mechanical designs \cite{fisicaroActiveStabilization2024,pallmannHighlyStable2023} have enabled lower vibration figures, allowing next-generation cavity platforms to achieve higher experimental duty cycles and vibration-averaged coupling rates. Ultimately, by combining strong cavity coupling with strain engineering techniques to prolong spin lifetimes, GeV defects offer a path towards a coherent spin-photon interface at liquid helium temperatures.

\section*{Acknowledgements}
Thanks to Thomas Clark for assisting with assembly of the fiber cavity.
This work is partially supported by the National Science and Engineering Research Council (NSERC RGPIN-2020-04095), Canada Research Chairs (229003), Fonds de Recherche – Nature et Technologies (FRQNT PR-247930), the Canada Foundation for Innovation (Innovation Fund 2015 project 33488 and LOF/CRC 229003), and l’Institut Transdisciplinaire d’Information Quantique (INTRIQ).

\pagebreak
\appendix

\section{Diamond Sample Fabrication}
\label{sec:appendix_sample_fabrication}
The diamond sample starts as an electronic grade $\langle 100 \rangle$-cut diamond (Element 6), which is laser sliced (Delaware Diamond Knives) and ion implanted at a fluence of $10^9$ Ge/\SI{}{\per\cm\squared} with an energy of $\SI{330}{KeV}$ (Innovion), targeting a depth of $\SI{125\pm27}{nm}$ (calculated with SRIM \cite{zieglerSRIMStopping2010}). Following implantation, the sample is annealed with a three step process \cite{chuCoherentOptical2014},  yielding a typical density of $<1$ GeV/\SI{}{/\micron\squared}~\cite{hoyjensenCavityEnhancedPhotonEmission2020}. Subsequently, the sample is thinned by reactive ion etching (RIE) with inductively coupled plasma (ICP) \cite{rufOpticallyCoherentNitrogenVacancy2019} to the desired thickness ($\sim \SI{1}{um}$). After the etch, the sample is acid cleaned with a 2:1 piranha solution and deposited face-down on the mirror where it bonds via intermolecular forces \cite{hoyjensenCavityEnhancedPhotonEmission2020}.

\section{Cavity Vibrations}
\begin{figure}[bh]
    \includegraphics{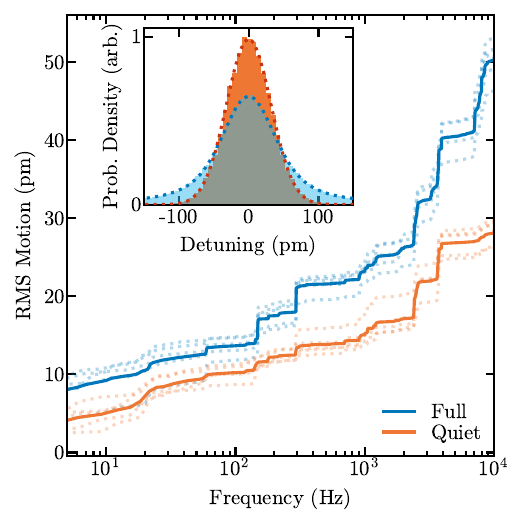}
    \caption{\label{fig:vibrations} Vibrations Characterization. Cumulative root-mean-square cavity motion as a function of frequency during the full cryo cycle (blue) and quiet portion (orange). Solid lines are averaged from individual dotted measurements. Inset shows density histogram of detunings, demonstrating the quiet portion's smaller range of deviations. Dotted blue (red) line shows Voigt (Gaussian) fit to full (quiet) data, from which vibration contributions are determined, as presented in text.}
\end{figure}

To measure deviations in the cavity throughout the compressor cycle we adopt a technique similar to that presented in \cite{ruelleTunableFiber2022}.
As in the measurement of the cavity linewidth presented in the main text, we phase modulate the \SI{602}{\nano\meter} beam sent to the cavity at 4 GHz and collect the free-space transmission while sinusoidally driving the cavity fiber mirror at \SI{10}{\kilo\hertz}. Since each oscillation cycle sweeps through resonance twice, the measurement rate is twice the sweep frequency, and we analyze the up- and down-sweep data separately to eliminate piezo hysteresis effects.
By measuring the deviation of each resonance from the mean, calibrated to the known side-band frequency, we determine the detuning in frequency induced by vibrational motion or drift of the cavity mirrors.
Using the known mode-structure slope we can convert frequency detuning to length detuning for each observed resonance. Data from the quiet periods are then post selected based on a signal synchronized to the cryopump.
To determine the rms motion, we compute cumulative integrals of the PSD derived from the measured length deviations, as presented in \cref{fig:vibrations}.
Averaging over a few acquisitions, we determine the full (quiet) motion to be \SI{51\pm1}{\pico\meter}-rms (\SI{28.9\pm0.7}{\pico\meter}-rms) 
\label{sec:appendix_vibration}
While this technique is straightforward to implement, it interprets sweep-to-sweep variations in piezo response as vibrations, potentially resulting in an additional contribution not present in the undriven cavity.

In order to characterize the distribution of the vibrations we compute histograms of the detunings as shown in the inset of \cref{fig:vibrations}.
We find the quiet period distribution to be well fit by a Gaussian with $\sigma = \SI{31.7\pm0.1}{\pico\meter}$. 
The full distribution has long tails at large detunings and is well fit by a Voigt distribution yielding a Lorentzian FWHM ($\Delta_{\rm{L}}$) of $\SI{43.7\pm0.5}{\pico\meter}$ when fixed to the same Gaussian width as the quiet histogram.
Converted to frequencies, these correspond to \ghzpi{\sigma_{\nu}} {1.796\pm0.006} and \ghzpi{\Delta_{\nu}}{2.48 \pm 0.03}.

\section{Intra-cavity Spectroscopy}
\label{sec:appendix_intracav}
For experiments probing the linewidth of the cavity-coupled GeV emission by scanning the cavity length, we need to account for vibrations in interpreting observed spectra. 
Since we perform these experiments without synchronization to the cryostat cycle, the full vibration distribution contributes.
To account for potential variations in vibration amplitude, particularly during the noisy periods, we simultaneously pass a near-resonant, narrow-linewidth (< MHz) laser through the cavity as an \textit{in situ} reference beam with known linewidth. We fit both the emitter and laser peaks to Voigt profiles constrained to have Gaussian linewidth $\sigma_{\nu}$, and extract their respective Lorentzian linewidths $\Delta_e$ and $\Delta_l$. 
$\Delta_{l}$ comprises the sum of the intrinsic cavity linewidth $\kappa'$ and the additional Lorentzian contribution to vibrations present during data acquisition.
Thus the final emitter linewidth of interest, with vibrations subtracted, is $\Delta_{\rm{cw}}=\Delta_{e} - (\Delta_{l} - \kappa')$. 
Statistical error bars are obtained by averaging the results obtained over several scans.
As a check on our procedure for estimating the vibrational contribution to the emitter spectrum, we can compare the independently measured Lorentzian component of the vibrations 
\ghzpi{\Delta_{\nu}}{2.48 \pm 0.03} to the Lorentzian component inferred from the laser peak in our spectra, \ghzpi{(\Delta_{l} - \kappa')}{2.3\pm 0.1} indicating a small variation of the full cycle vibration amplitude.

To accurately interpret the data in \cref{fig:emitter}a, we needed to remeasure 
the cavity linewidth $\kappa'/2\pi$, which had increased to \SI{5.8\pm 0.1}{\giga\hertz} 
from the value of \SI{2.07 \pm 0.06}{\giga\hertz} that we measured earlier 
during data acquisition for lifetime measurements. The spectra similar to those shown in \cref{fig:emitter}a are the only data sets for which this higher linewidth applies, as they were the final acquisition. Note that the increase in the cavity decay rate does not affect other cQED parameters (in particular $g$ is set by the cavity mode volume and emitter characteristics), allowing us to combine measurements taken at different finesse values in determining $g$.  

This degradation in finesse occurred following extended exposure to a high-power ps-pulsed (>\SI{0.10}{\nano\joule}, \SI{40}{\mega\hertz}) laser while acquiring data shown in \cref{fig:lifetime}e,f. Such degradation was also observed for high-power (>\SI{30}{\milli\watt}) CW pump exposure on another defect. It was not associated with any degradation in the images produced by our system (we have observed such degradation following vacuum issues leading to ice deposition).
Damage was observed following high-power exposure at several positions, indicating that it was likely not due to local contamination.
Additionally, the increase in cavity absorption was localized to small ($\sim$\SI{1}{\micron\squared}) regions.
These behaviours suggest that the degradation may stem from laser-induced damage to either the mirror coating or diamond surface.
\section{Weak Coupling cQED}
\label{sec:appendix_cqed}
To model the lifetime reduction in our system we start from the usual Jaynes-Cummings (J-C) Hamiltonian
\begin{align*}
    \mathcal{H} = \hbar \nu a^\dagger a + \hbar \omega \sigma_{ee} - \hbar g (a \sigma_{eg} - a^\dagger \sigma_{ge}),
\end{align*}
where $\nu$ is the cavity frequency, $\omega$ the emitter transition frequency, and $g$ the cavity coupling rate \cite{jaynesComparisonQuantum1963,shoreJaynesCummingsModel1993}. We consider a two-level emitter with a ground $(\ket{g})$ and excited $(\ket{e})$ state, coupled by the raising (lowering) operator $\sigma_{eg}$ $(\sigma_{ge})$. Since we consider an initial state where the emitter is excited and the cavity is empty, the cavity is similarly restricted to either the $\ket{0}$ or $\ket{1}$ Fock state.
To model the incoherent effects, we use the Lindblad master equation
\begin{align}
    \label{eq:master}
    \dot{\rho} = \frac{1}{i\hbar}\comm{\mathcal{H}}{\rho} - \sum_k\frac{\zeta_k}{2}\left(\left\{L_k^\dagger L_k,\rho\right\}-2L_k\rho L_k^\dagger\right), 
\end{align}
where $L_k$ represents the decoherence jump operators occurring at rate $\zeta_k$. In this case, the decoherence effects considered are spontaneous emission occurring at rate $\gamma$ with jump operator $\sigma_{ge}$, pure dephasing at rate $\gamma^*$ with jump operator $\sigma_{ee} = \sigma_{eg}\sigma_{ge}$ and cavity loss at rate $\kappa$ with jump operator $a$.

We did not observe blinking or strong spectral jumps that would be consistent with other sources of noise and decoherence, such as charge noise, and the observed broadening of the GeV emission lines relative to cavity-coupled laser transmission could be accounted for by a purely Lorentzian contribution to the linewidth. We therefore model the dephasing within a master equation formalism, as we see no evidence of more complex noise processes. 

We solve for the populations in the excited atom $(\rho_{aa})$ and occupied cavity $(\rho_{cc})$ states by adiabatic elimination of the coherences $\dot{\rho}_{ac} = 0$ obtaining

\begin{align*}
    \rho_{aa}(t) &= \frac{1}{2\alpha}e^{-\frac{t}{2}\left(2R+\alpha+\gamma+\kappa\right)}\left(\alpha + \gamma - \kappa + e^{\alpha t}\left(\alpha-\gamma+\kappa\right)\right),\\
    \rho_{cc}(t) &= \frac{R}{\alpha}e^{-\frac{t}{2}\left(2R+\alpha+\gamma+\kappa\right)}\left(e^{\alpha t} -1\right),
\end{align*}
where
\begin{align*}
    \alpha &= \sqrt{4R^2 + (\gamma - \kappa)^2},\\
    R &= 4g^2\frac{\left(\kappa + \gamma + \gamma^* \right)}{\left(\kappa + \gamma + \gamma^* \right)^2 + 4\delta^2}=\frac{4g^2}{\Gamma}\frac{\Gamma^2}{\Gamma^2+4\delta^2},
\end{align*}
with a total decoherence rate $\Gamma = \kappa+\gamma+\gamma^*$.

Integrating these populations multiplied by their respective decay rates, we can determine the total emission probability from the atom and the cavity to respectively be
\begin{align*}
    p_{a} = \gamma \int_0^\infty \rho_{aa}(t)\dd{t} &= \frac{\gamma (R+\kappa)}{\gamma\kappa + R(\gamma + \kappa)},\\
    p_{c} = \kappa \int_0^\infty \rho_{cc}(t)\dd{t} &= \frac{\kappa R}{\gamma\kappa + R(\gamma + \kappa)}.
\end{align*}
Assuming these probabilities occur due to a competition of rates between the bare emission rate $\gamma$ and the new rate of emission via the cavity $\gamma_{\rm{cav}}$ we can write
\begin{align*}
    p_a = \frac{\gamma}{\gamma+\gamma_{\rm{cav}}},
    p_c = \frac{\gamma_{\rm{cav}}}{\gamma+\gamma_{\rm{cav}}},
\end{align*}
from which we determine
\begin{align*}
    \gamma_{\rm{cav}} = \frac{R\kappa}{R+\kappa} &=  \frac{4 g^2\Gamma}{\Gamma^2+\frac{4 g^2\Gamma}{\kappa}+4\delta^2},
\end{align*}
resulting in a total Purcell enhanced decay rate $\gamma_{\rm{P}} = \gamma + \gamma_{\rm{cav}} = \gamma + \frac{R\kappa}{R+\kappa}$. Defining $\Delta = \sqrt{\Gamma^2 + 4g^2\Gamma/\kappa}$ and $\tau_P = 1/ \gamma_P$ leads directly to \cref{eq:tauratio}. 
Note that in the fast-cavity limit $\kappa \gg R$, $\gamma_{\rm{P}}
= \gamma + R$, as is usually considered.




\section{Incoherently Driven cQED}
\label{sec:appendix_driven}
 Continuously driving the emitter affects the linewidth of the cavity-coupled emitter spectrum.  To account for it, we add an additional incoherent term to the J-C master equation \cref{eq:master}, with rate $\Pi$ and jump operator $\sigma_{eg}$, corresponding to incoherent excitation of the emitter.
The inclusion of this driving operator can couple the system to an infinite set of states with any number of photons in the cavity \cite{agarwalSteadyStatesCavity1990}. However, in the case of weak coupling $(g \ll \kappa,\gamma,\gamma^*)$, cavity photons are lost much faster than they are generated, so we truncate the basis to at most one photon in the cavity, allowing for four total states: $\ket{g} = \ket{g,0}$, $\ket{a} = \ket{e,0}$, $\ket{c} = \ket{g,1}$, $\ket{b} = \ket{e,1}$.

While the full time dynamics of the populations cannot be readily expressed analytically, we obtain steady-state populations by setting $\dot{\rho} = 0$, and determine the expected cavity occupation
\begin{align*}
    \expval{n} &= \expval{\rho}{c}+\expval{\rho}{b}\\
    &= \frac{4g^2\Pi\,\Gamma_{\rm{cw}}}{4g^2\Gamma_{\rm{cw}}(\gamma+\kappa+\Pi) + \kappa(\gamma+\Pi)(4\delta^2 + \Gamma_{\rm{cw}}^2)},
\end{align*}
where $\Gamma_{\rm{cw}} = (\gamma+\gamma^*+\kappa+\Pi)$. Considering the measured values presented in the main text, we calculate the occupation of the doubly excited $\ket{b}$ state to be $\rho_{bb} = (2\pm1)\cdot10^{-4}$, confirming the validity of neglecting any higher photon-number states.

Since both the cavity $|c\rangle$ and doubly excited state $|b\rangle$ will emit photons from the cavity at rate $\kappa$, we can determine the cavity-coupled fluorescence rate as
\begin{align*}
    \kappa\langle n\rangle &= A_{\rm{cw}}\frac{\Delta_{\rm{cw}}^2}{\Delta_{\rm{cw}}^2+4\delta^2},\\
    A_{\rm{cw}} &= \frac{4 \Pi  g^2\kappa}{\kappa  (\gamma +\Pi ) \Gamma_{\rm{cw}}+4 g^2 (\gamma +\Pi +\kappa )}\\
    \Delta_{\rm{cw}} &= \sqrt{\Gamma_{\rm{cw}} \left(\Gamma_{\rm{cw}}+\frac{4 g^2 (\gamma +\Pi +\kappa )}{\kappa 
   (\gamma +\Pi )}\right)}.
\end{align*}
The functional form indicates that as we sweep the cavity over the emitter resonance, the collected intensity follows a Lorentzian profile with a linewidth given by $\Delta_{\rm{cw}}$, additionally broadened by the cavity coupling from the expected total linewidth $\Gamma_{\rm{cw}}$.

Similarly, we determine the coupled system saturation behaviour by analyzing the amplitude of the emitted fluorescence $A_{\rm{cw}}$. At low pump rates, the denominator is dominated by the linear term. With the goal of reaching a functional form of a saturation curve ($I_{\rm{max}}P/(P+P_{\rm{sat}})$), we take the first order Pad\'e approximant in terms of pump rate
\begin{align*}
    A_{\rm{cw}}&\approx A_{\rm{cw}}^{(1,1)}\\
    &= \frac{4g^2\kappa}{4g^2+\kappa(2\gamma+\gamma^*+\kappa)}\frac{\Pi}{\gamma+\frac{(4g^2-\gamma^2)\kappa}{4g^2+\kappa(2\gamma+\gamma^*+\kappa)}+\Pi}
\end{align*}
From which we determine
$$\Pi_{\rm{sat}} = \gamma + \frac{(4g^2-\gamma^2)\kappa}{4g^2+\kappa(2\gamma+\gamma^*+\kappa)}$$
As in our system, if we consider $\gamma \ll g, \kappa,\gamma^*$ then we can approximate
$$\Pi_{\rm{sat}} \approx \gamma + \frac{4g^2\kappa}{4g^2+\kappa\Gamma}, $$
identical to the on-resonance cavity-coupled decay rate.

\section{Bounding quantum efficiency}
\label{sec:appendix_estimatingQE}

The coupling rate between the cavity and an ideal emitter in the diamond membrane,  $g_0 = |E_{\rm{max}}||\mu|/\hbar$, can be found by determining the emitter dipole moment $\mu$ and the electric field maximum $E_{\rm{max}}$ in the diamond for a single photon in the cavity. For an ideal two-level emitter with lifetime $\tau$, 
\begin{align}
    \mu^2 = \frac{3 \epsilon_0 \lambda^3 \hbar}{8 n \pi^2 \tau},
\end{align}
where $\lambda$ is the resonant wavelength in vacuum, $\epsilon_0$ is the permittivity of free space, and $n$ is the index of refraction in which the emitter is located. For a resonant wavelength of 602 nm and a $\SI{6.1\pm0.1}{\nano\second}$ lifetime, this yields a dipole moment of 2.3$\times 10^{-29}$ C-m. 

The maximum single-photon electric field can be found by appropriately quantizing the intra-cavity electromagnetic energy \cite{riedelCavityEnhancedRamanScattering2020}:
\begin{align}
    \int_{\text{cavity}} \epsilon_0 n^2(z) |\mathbf{E}(x,y,z)|^2 dV = \frac{\hbar \omega}{2},
\end{align}
where $\omega = 2 \pi c/\lambda$ and $n(z)$ indicates the position-dependent index of refraction. We calculate the intra-cavity energy numerically: we assume a Gaussian waist equal to the experimentally measured value (as in \cref{fig:lifetime}c), $\SI{1.30\pm0.01}{\micron}$ (GeV 1,2), $\SI{1.40\pm0.01}{\micron}$ (GeV 3) and model the longitudinal mode via a 1D transfer-matrix model that uses the nominal mirror coating layers, with the diamond membrane $\SI{1.05\pm0.02}{\micron}$ (GeV 1,2) and $\SI{0.79\pm0.02}{\micron}$ (GeV 3) and air-gap thicknesses $\SI{6.95\pm0.05}{\micron}$ (GeV 1), $\SI{6.43\pm0.05}{\micron}$ (GeV 2) and $\SI{9.71\pm0.03}{\micron}$ (GeV 3) extracted from fitting cavity transmission spectra (as in \cref{fig:setup}d). Setting the intra-cavity energy equal to $\hbar\omega/2$ 
allows extraction of the electric field maximum. In combination with $\mu$ given above, we can thereby calculate $g_0$ for the cavity parameters (waist, air and diamond membrane thicknesses) appropriate to each of our emitters. 


For non-ideal emitters, the coupling rate $g$ is related to the ideal coupling rate by $g = \sqrt{\eta} g_0$ with $\eta = \eta_{QE}\,\eta_{DW} \,\eta_{BR}\,\eta_{Z} \cos^2\alpha$ (see main text for definitions). As discussed in the main text, $\eta_{\rm{DW}} = 0.6$ allows us to place a strict lower bound on $\eta_{\rm{QE}}$; here, we consider likely values for the other factors, and whether our observations are consistent with the possibility of unity quantum efficiency. Although the orientation of the GeV dipole moment is not known, it's expected to have similar properties to other group-IV color centers \cite{nguyenPhotodynamicsQuantum2019}. For SiV centers it has been shown that their ZPL is predominately polarized along the $\langle 111 \rangle$ direction with a perpendicular component in the $(111)$ plane \cite{heppElectronicStructure2014,rogersElectronicStructure2014}. We calculate both cases where a $\langle 111 \rangle$-oriented dipole would produce $\cos^2{\alpha} = 2/3$ and a dipole moment averaged over the $(111)$ plane corresponds to $\cos^2{\alpha} = 1/2$ and take the more conservative and likely value in our estimate in the main text. Next, $\eta_{BR} =  P_{(1-3)}/(P_{(1-3)} + P_{(2-3)})$ where $P_{(x-y)}$ is the probability of emission on the $x-y$ transition. $\eta_{BR}$ cannot be reliably determined from our cavity spectroscopy data, as the observed heights of the emission peaks are influenced by intra-branch transitions and cavity coupling (which depends on dipole orientation), but using the peak height ratio as an imperfect estimate suggests $\eta_{\rm{BR}}\sim 0.7-0.9$, around the same value $\eta_{\rm{BR}} = 0.77 \pm 0.08$ reported for SiVs \cite{mullerOpticalSignatures2014}. 
Finally, for the implantation parameters of our sample, the depth straggle is reasonably well approximated by a Gaussian distribution with standard deviation $27.1$~nm (SRIM \cite{zieglerSRIMStopping2010}). Since the electric field for an emitter displaced by $z$ from the optimal depth is $|E(z)| = |E_{\rm{max}}|\cos{(2\pi n z/\lambda)}$, a displacement of $\sigma = 27.1$~nm corresponds to  $\eta_{Z} = 0.6$.  For every emitter, the product of these reasonable values can be smaller than the calculated bounds on $\eta_{\rm{QE}}$. In particular, assuming $\eta_{DW} = 0.6, \cos^2{\alpha} = 2/3,$ and $ \eta_{BR} = 0.77 $, a straggle of $z = $ $\SI{21}{\nano\meter}$, $\SI{36}{\nano\meter}$, $\SI{25}{\nano\meter}$  would correspond to unity quantum efficiency in GeVs 1, 2, and 3 respectively. Thus, while it is likely that $z$ is small since we select emitters based on brightness, and it is possible that the dipole orientation could be influenced by strain to obtain larger $\cos^2{\alpha}$, the range of reasonable parameters certainly includes the possibility of unity quantum efficiency. Consequently, we cannot constrain the quantum efficiency from above. 
 
\bibliography{gev_cav}
\end{document}